\documentclass{vgtc}                          

\graphicspath{{figures/}{./}}

\usepackage{times}

\usepackage{mathptmx}
\usepackage{url}

\onlineid{0}
\vgtccategory{Research}
\vgtcinsertpkg

\title{Santiago's A.T. Field: Visualizing Urban Accessibility through an
\textit{Evangelion}-Inspired Interface}

\author{Eduardo Graells-Garrido\\
        \scriptsize CENIA \& Dept. of Computer Science\\
        \scriptsize Universidad de Chile
\and Ignacio P\'erez-Messina\\
     \scriptsize TU Wien
\and Claudio Gaete\\
     \scriptsize Dept. of Computer Science\\
     \scriptsize Universidad de Chile}

\teaser{
  \centering
  \includegraphics[width=.4\linewidth]{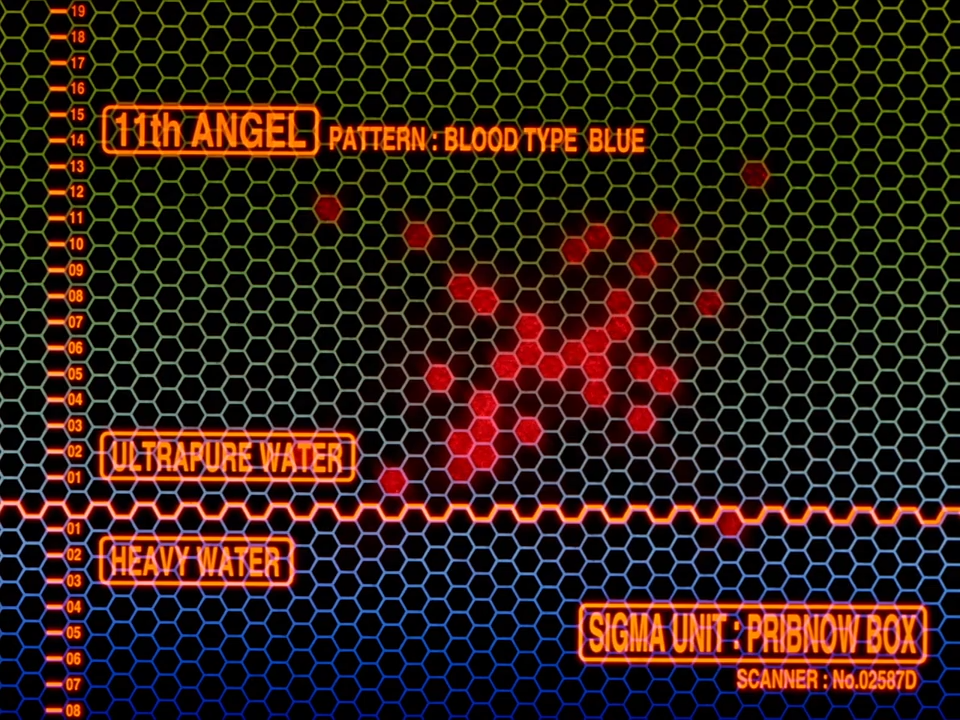}\includegraphics[width=.6\linewidth]{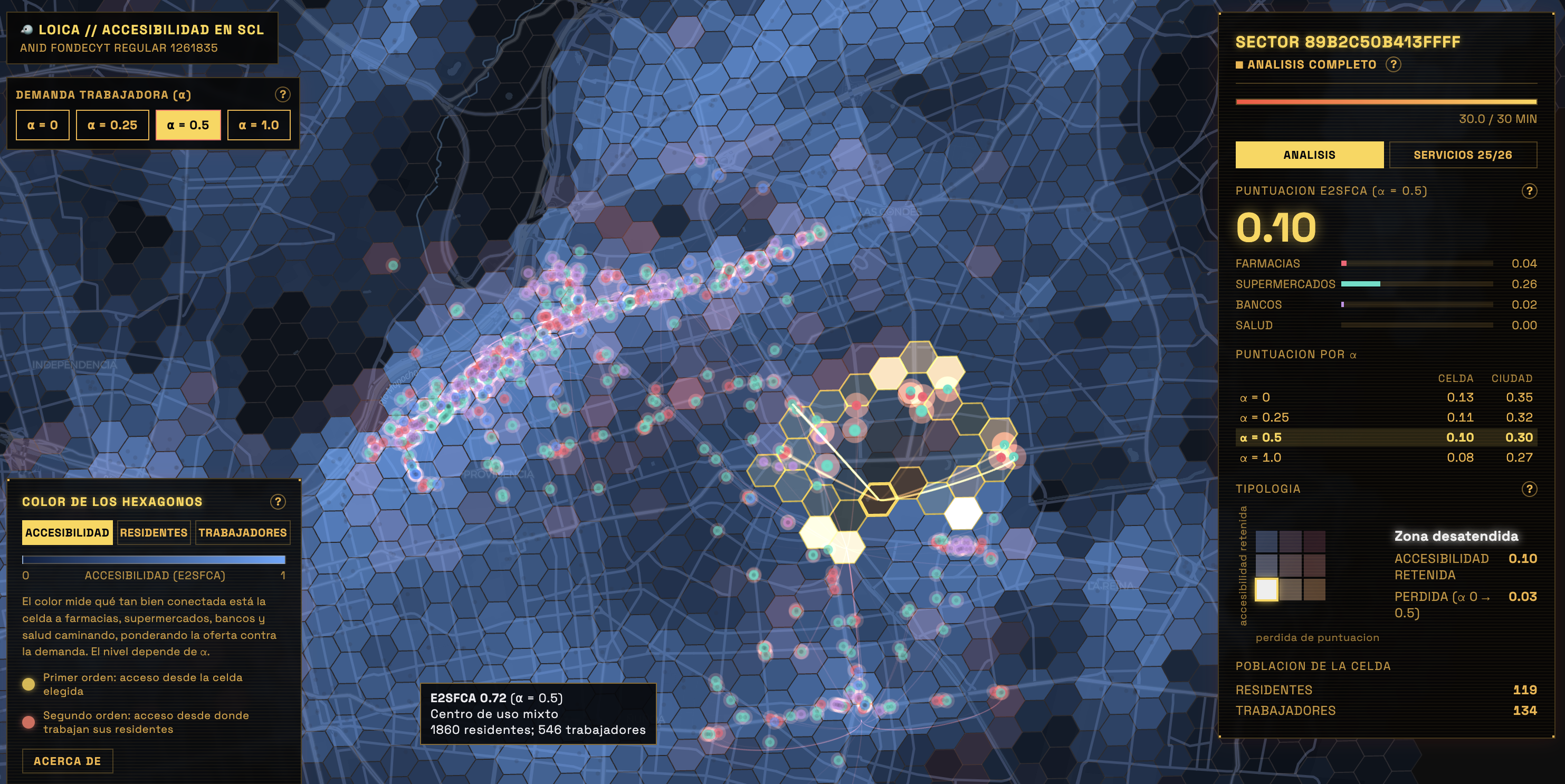}

  \caption{A fictional display and its translation into a working
  visualization.
  Left: a frame of Episode 13 of \textit{Neon Genesis
  Evangelion}~\cite{anno1995}, with the interface used to identify the Angel
  Ireul. Right: Santiago's A.T.
  Field, an interactive visualization of urban accessibility.}
  \label{fig:translation}
}

\abstract{
Science-fiction interfaces are often reproduced for their appearance, and their
diagnostic logic is rarely rebuilt around real data. Here we describe
Santiago's A.T. Field, an interactive visualization of urban accessibility
built from a diagnostic interface in Episode 13 of \textit{Neon Genesis
Evangelion}. We translated that interface in five stages: phenomenon, field,
progression, interface, and response. A microscopic Angel became a relational
urban problem, and a fictional scanning lattice became an accessibility field
computed with the E2SFCA method. Every element of the fictional display is tied
to a quantity of the model. Three tensions appeared when we made the fictional
language operational: the saturated palette of the genre against color-vision
deficiency, the motion of the genre against the reduced-motion preference, and
the per-frame cost of the animation. The urgency of the fictional command
center has its own cost, because emergency aesthetics can cast urban
populations as threats. These three tensions are what the translation cost, and
they are what transfers to other attempts of this kind.
}


\newcommand{\episodequote}[2]{%
  \begin{quote}
    \small\itshape ``#1''\\[-0.2em]
    \hfill---#2
  \end{quote}
}

\begin{document}

\firstsection{Preface}
\maketitle

This paper answers SciFi-VIS's invitation to examine the two-way relation
between science fiction and visualization~\cite{gunther2026scifivis} through
one concrete act of translation. \textit{Neon Genesis Evangelion} is a 1995
Japanese animated series in which the organization NERV uses dense diagnostic
interfaces to interpret attacks by entities called Angels. Our visualization
began with one of those instruments. We asked what had to be modified for it to
operate on a real problem, through a process of \emph{constructive
translation}, which means rebuilding a fictional visualization around a new
analytical object. The five sections that follow are the five stages of that
work, and a last section states what it cost. The interactive
system is available~\cite{loicaSystem}.

Research on science-fiction interfaces has surveyed how films depict
interaction~\cite{schmitz2008survey,shedroff2012makeitso}, and it has described
fictional devices as prototypes of real technology~\cite{kirby2010diegetic}. In
visualization, the superpowers of fiction have been used to characterize and
inspire systems~\cite{willett2022superpowers}. Planning research has reviewed
accessibility instruments, which are tools to compute and display accessibility
for practitioners~\cite{papa2016accessibility}.

Studies of \textit{Evangelion} have analyzed its visual
style~\cite{santiago2021animestudies}, but we found no study of its interfaces.
The series belongs to a wider circulation of anime outside Japan. In Latin
America, comic artists have appropriated manga aesthetics and adapted them to
their local contexts~\cite{gutierrez2017manga}. In a similar way, we adopted the
interface aesthetics of \textit{Evangelion} and adapted them to an
accessibility instrument for Santiago.

\begin{figure*}[ht!]
  \centering
  \includegraphics[width=\linewidth]{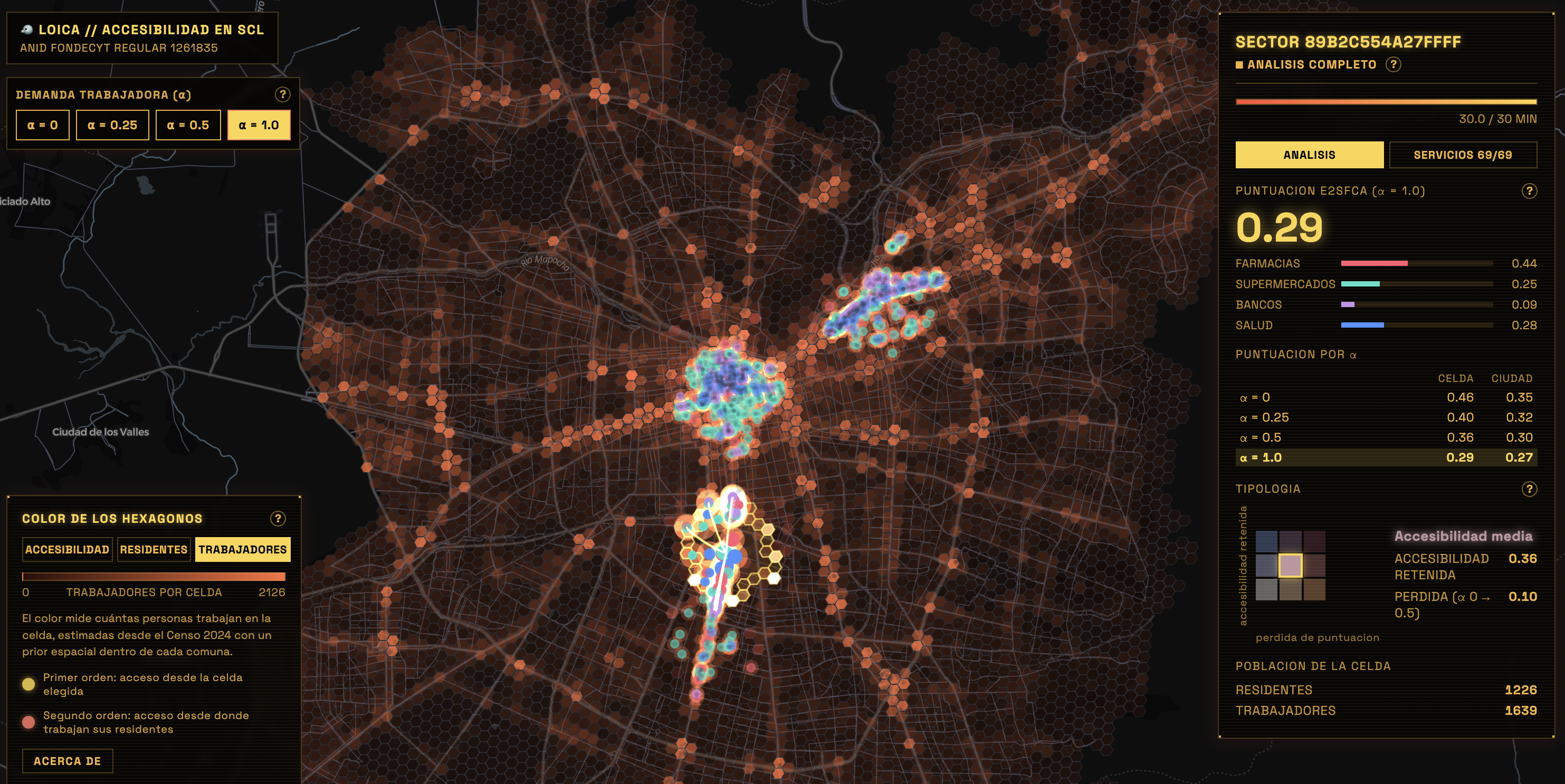}
  \caption{An intermediate state of Santiago's A.T. Field. The scan advances
  over the city and activates the services and the spatial relations that build
  the accessibility index.}
  \label{fig:progression}
\end{figure*}

\section{Phenomenon}

\episodequote{It's an Angel.}{Ritsuko, Episode 13~\cite{evawiki13}}

\noindent In Episode 13 of \textit{Neon Genesis Evangelion}, technicians find what looks
like corrosion on a wall inside NERV. It spreads and resists removal, and its
distributed behavior becomes visible only through the visualization interface,
where it grows across a hexagonal lattice. It is Ireul, the 11th
Angel~\cite{anno1995}.

We translate this to urban accessibility in Santiago de Chile, which is also a
distributed relation among people, services, and movement. It appears in one
place and is produced by the whole system. The two objects also differ. Ireul
is a hostile entity, and urban accessibility describes the opportunities and
exclusions of real inhabitants.

\section{Field}

\episodequote{The pattern analysis is blue.}{Ritsuko, Episode 13~\cite{evawiki13}}

\noindent Santiago is tessellated into 9,025 hexagonal cells of 0.09 km$^2$ with
the H3 grid at resolution 9. The H3 grid is built from hexagons because the center of a
cell is at the same distance from the centers of its six neighbors.
Each cell receives an accessibility value. The prototype
considers walking access to pharmacies, supermarkets, banks, and health
services. Using the Enhanced Two-Step Floating Catchment Area method (E2SFCA),
each service first receives a supply-to-demand ratio within a distance-weighted
catchment; a cell's accessibility then accumulates the ratios of the services
it can reach~\cite{luo2009e2sfca}.

Our implementation extends this relation through the city's residence--work
structure. The demand of a cell is its residents plus $\alpha$ times its
workers, and the interface offers four values of $\alpha$: 0, 0.25, 0.5 and 1.
With $\alpha = 0$ the index is the residential E2SFCA, and with $\alpha = 1$ a
worker competes for the services around the workplace as much as a resident.
The right value depends on how often workers use services near their
workplace, and our data do not measure it. For this reason, $\alpha$ remains a
parameter, and the interface opens at the midpoint, $\alpha = 0.5$.
We call the result commuter-adjusted accessibility. Selecting a cell shows what
its residents reach from home and what they reach from the places where they
work, so the value of one hexagon depends on the rest of the city.

This computed surface is Santiago's A.T. Field. In \textit{Evangelion}, A.T.
stands for \textit{Absolute Terror}, and the field is a barrier that blocks hostile contact.
Here it stands for \textit{Accesibilidad y Trabajo}, Spanish for accessibility
and work. Both displays use the same hexagonal form
(\autoref{fig:translation}). The fictional field blocks contact and this one
measures it.

\section{Progression}

\episodequote{Evolution\ldots}{Fuyutsuki, Episode 13~\cite{evawiki13}}

\noindent The interface scans the city, locates services, and activates the relations
through which the index is assembled (\autoref{fig:progression}). Every element
of the scan is tied to a quantity of the model. The front of the pulse advances
over walking times computed with R5~\cite{conway2017r5} over the OpenStreetMap
network, so its shape follows the street network. Walking times were computed at 3.6 km/h and end at 30 minutes. The catchment weight falls from 1 below 10
minutes to 0.03 above 20, so services beyond a 20-minute walk barely add to the
index. Each service lights up when the front reaches
it, and its size and brightness follow the weight of the E2SFCA catchment, so a
service that adds little to the index is drawn small and dim. Gold arcs join
the cell to the services reached from home, and red arcs join it to the
services reached from the workplaces of its residents.

Progression exposes how services, catchments, and demand accumulate. The model describes cumulative reachability, and the front of the scan marks the walking time from the selected cell.

\section{Interface}

\episodequote{These are back doors!}{Maya, Episode 13~\cite{evawiki13}}

\noindent Both interfaces overlay a hexagonal lattice on a dark field, with cool context,
warm active regions, and luminous technical labels. The result reads as an
institutional diagnostic system.

In the fictional display, these elements need only sustain narrative
plausibility. In the real interface, they must carry reproducible meanings. The
grid encodes four variables and each one has its own ramp: blue for
accessibility, gold for residents, orange for workers, and a bivariate
three-by-three palette that crosses retained accessibility against the share of
the daytime population which arrives to work.

The data were prepared with the following Python libraries: tobler for the
interpolation of census counts to the grid, QuackOSM for the extraction of
pharmacies, supermarkets and banks from OpenStreetMap, and
r5py~\cite{fink2022r5py} for the walking times. The interface was implemented
with React, deck.gl for the map layers, MapLibre GL for
the basemap and h3-js for the cell geometry.

Three differences appeared when we made the fictional
interfaces into a real one. First, the dark interface asks for saturated color, and
the bivariate palette has to stay separable under color-vision deficiency. We
built that palette in CIELCh space for a printed figure, and we validated its
nine opaque colors with a physiological simulation of the three
deficiencies~\cite{machado2009physiologically}. The interface lightens those colors for the dark basemap and raises their opacity with the value. The validation does not cover these modified colors. Second, the genre
relies on motion, and some users set a reduced-motion preference in the
browser. With that preference, the assembled field appears without the scan. A
control in the legend also turns the scan on and off.
Third, the animation has a frame budget: we animate inside the color accessor
of each layer, because rebuilding layer data on every frame re-tessellates the
grid and the frame rate drops.

The fictional display claims authority before its measures are explained. Our
reconstruction keeps that authority, and the explanations and the parameters
work against it, because they expose the assumptions of the model.

\section{Response}

\episodequote{The end point of evolution is death.}{Gendo, Episode 13~\cite{evawiki13}}

\noindent Episode 13 stages visual analytics under emergency. An unfamiliar phenomenon
changes faster than the institution can understand it. Automated systems detect
its patterns but cannot respond alone, and Ritsuko interprets the adaptation and
intervenes. Visualization mediates detection and action under time pressure.

Urban accessibility is a \emph{slow emergency} that accumulates through
ordinary journeys and unequal infrastructure. Santiago's A.T. Field imports the
urgency of the fictional command center into this structural problem, and asks
where connection is insufficient.

Emergency aesthetics can make persistent inequality immediate, and they can
also cast urban populations as threats. We have not resolved this tension.

\section{Outlook}

The system covers Santiago, and it has not been used in planning practice yet.
A replication in another city needs three inputs: resident counts, an
OpenStreetMap street network and the location of jobs. The walking range could
also become a control. This change requires precomputing the index for each
range, in the same way as for the four values of $\alpha$.

The scan presents the index as a sequence, which relates it to narrative
visualization~\cite{segel2010narrative}. Studies of charts show that visual embellishment and other visual features
change what readers remember~\cite{bateman2010useful,borkin2013memorable}. A
study could compare the scan with a static map of the same index, and measure
what readers understand and remember from each one.

\section{Transfer}

Santiago's A.T. Field translates a chain of relations. The fictional scanning
lattice became an analytical substrate, and the dramatic animation became an
explanation of how the index is built. The three tensions of the Interface
section are the cost of the translation, and they are what transfers to other
attempts at making a science-fiction instrument operational.

\textit{Evangelion} was released in 1995 and set its story in 2015.
Santiago's A.T. Field carries one of its instruments into a working interface
in 2026.

\acknowledgments{
This work was supported by the Chilean Agency of Research and Development through ANID FONDECYT Regular 1261835 (project LOICA), through the
National Center for Artificial Intelligence CENIA FB210017, Basal ANID, and by the TU Wien.
}

\bibliographystyle{abbrv-doi}
\bibliography{template}

\end{document}